\documentclass{INTERSPEECH2023}

\interspeechcameraready

\usepackage{amsmath, graphicx, multirow, caption, amssymb, hyperref, pifont, soul}

\title{Exploring Train and Test-Time Augmentations for Audio-Language Learning}
\name{%
\begin{tabular}{@{}c@{}}
Eungbeom Kim$^{1\star}$, Jinhee Kim$^{2\star}$, Yoori Oh$^2$, Kyungsu Kim$^2$,\\ 
Minju Park$^2$, Jaeheon Sim$^1$, Jinwoo Lee$^2$, Kyogu Lee$^{1,2,3}$
\end{tabular}\thanks{$^\star$ Equal contribution.}}
\address{$^1$Interdisciplinary Program in Artificial Intelligence,
    $^2$Department of Intelligence and Information,\\
	$^3$Artificial Intelligence Institute, Seoul National University, Republic of Korea}
\email{eb.kim@snu.ac.kr}

\begin{document}

\maketitle

\begin{abstract}
In this paper, we aim to unveil the impact of data augmentation in audio-language multi-modal learning, which has not been explored despite its importance. We explore various augmentation methods at not only train-time but also test-time and find out that proper data augmentation can lead to substantial improvements. Specifically, applying our proposed audio-language paired augmentation PairMix, which is the first multi-modal audio-language augmentation method, outperforms the baselines for both automated audio captioning and audio-text retrieval tasks. To fully take advantage of data augmentation, we also present multi-level test-time augmentation (Multi-TTA) for the test-time. We successfully incorporate the two proposed methods and uni-modal augmentations and achieve 47.5 SPIDEr on audio captioning, which is an 18.2\% relative increase over the baseline. In audio-text retrieval, the proposed methods also show an improvement in performance as well.

\end{abstract}

\section{Introduction}
\label{sec:intro}

Data augmentation has shown meaningful performance gain in various domains. Multi-modal learning is one of the areas where data augmentation is prospective and various augmentation strategies are applicable. We can apply different augmentations on each modality such as audio augmentations \cite{park2019specaugment, wang2021specaugment++} and text augmentations \cite{wei2019eda, edunov2018understanding}. We can also augment multi-modalities jointly. MixGen \cite{hao2023mixgen}, for instance, is a simple way to generate image-text pairs by applying mixup \cite{zhang2018mixup} on images and concatenating each corresponding text.

However, data augmentation on audio-language learning remains poorly understood. Even worse, AudioCaps \cite{kim2019audiocaps}, the largest public audio-language dataset, only includes about 40K audio clips and 46K captions in opposition to the MS COCO Captions \cite{lin2014microsoft} which is a widely known vision-language dataset containing 330K images and 1.5M captions.
For this reason, we delve into data augmentation on audio-language learning in this study.

\begin{figure}
    \centering
    \includegraphics[width=1\columnwidth]{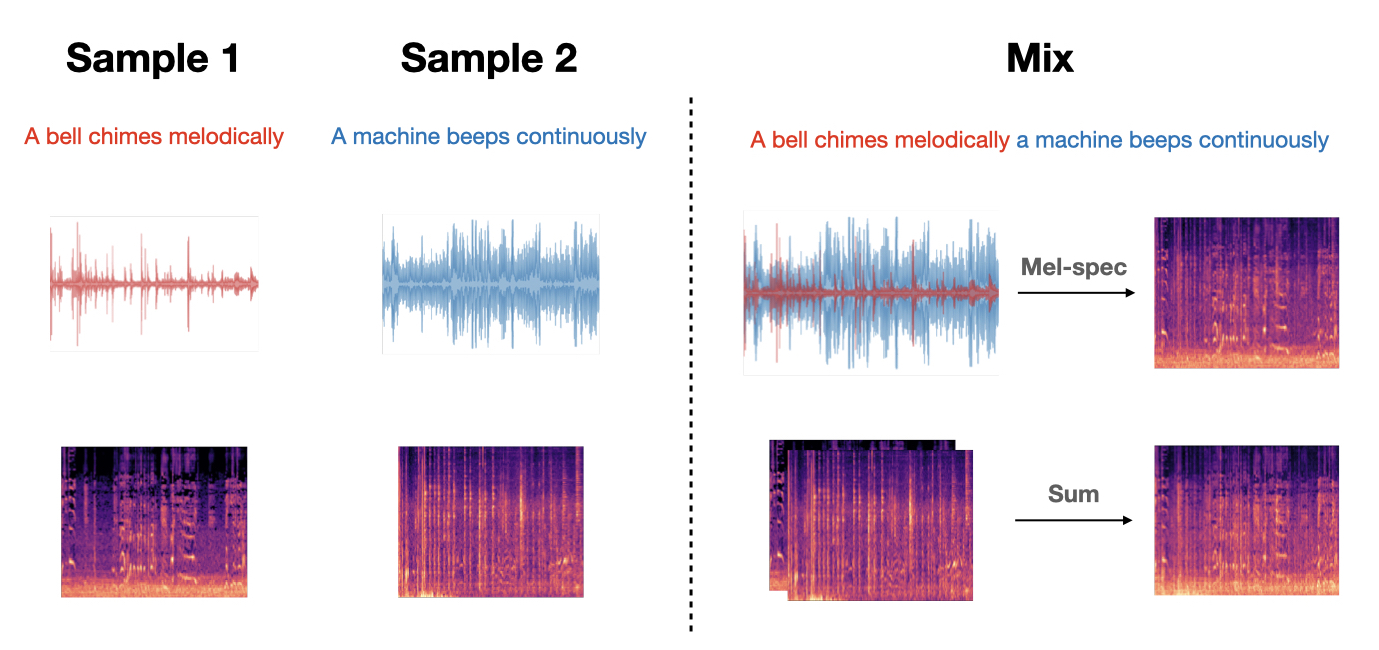}
    \caption{Illustration of PairMix. PairMix mixup audio clips and concatenates the corresponding text. Audio mixup can be implemented at waveform-level or mel spectrogram-level.} 
    \label{fig:mixgen}
\end{figure}

We first argue that paired multi-modal data augmentation takes an important role in audio-language learning. Therefore, as a train-time augmentation, we propose PairMix which is a simple and effective audio-text paired augmentation. 
PairMix generates audio-text pairs by randomly selecting either waveform-level or mel spectrogram-level mixup with a certain probability for each generated sample and concatenating their corresponding text captions, as illustrated in Figure \ref{fig:mixgen}. 
To support our argument, we first analyze the uni-modal augmentations for audio-language learning and the proposed multi-modal augmentation method, PairMix. As a result, we observe the importance of multi-modal augmentation in modeling the audio-language relationship. PairMix not only outperforms other data augmentation methods but also helps uni-modal data augmentation when uni-modal augmentation and PairMix are applied at the same time. 
Since the audio modality has different design choices compared to the image, we also compared PairMix with other audio augmentation design choices. PairMix surpasses other designs for multi-modal augmentation such as waveform-level mixup, mel spectrogram-level mixup, and audio concatenation.

Test-time augmentation (TTA) is another data augmentation approach that is shown to increase performances across diverse tasks \cite{moshkov2020test, molchanov2020greedy}. However, TTA is disregarded in most audio-language learning studies. In this paper, we investigate the effect of TTA on audio-language learning. The problem with TTA is that it has bounded performance, and fail to scale up with the size of augmentation.
To address this issue, we approach TTA from a new perspective, the place of aggregation. We present a novel multi-level test-time augmentation (Multi-TTA) methodology, which is a generalization of TTA over multiple layers. Multi-TTA successfully outperforms the baselines and we discover that there is still room for improvement using augmentations at test-time.

\begin{figure*}
    \centering
    \includegraphics[width=1\textwidth]{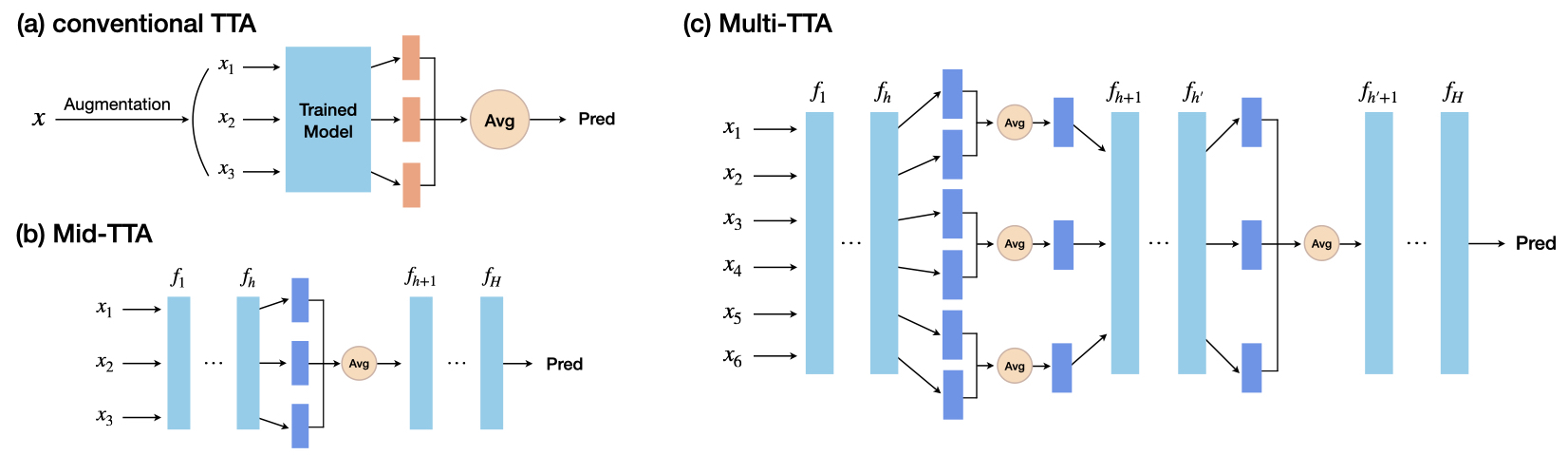}
    \caption{Frameworks for test-time augmentation (TTA). With the augmented inputs at test-time, (a) Conventional TTA averages the final predictions, (b) Mid-TTA averages the intermediate representations, and (c) Multi-TTA averages the outputs of multiple layers at different levels.}
    \label{fig:tta}
\end{figure*}

We evaluate the proposed methods across automated audio captioning (AAC) and audio-text retrieval on AudioCaps dataset. In AAC, PairMix with audio augmentations achieves 46.6 SPIDEr, which is a 15.9\% over the baseline without modifying its architecture, and outperforms the state-of-the-art method with a 5x smaller model. Applying PairMix and Multi-TTA results in an extra improvement and attains 47.5 SPIDEr. PairMix and Multi-TTA are also effective in audio-text retrieval. Compared to the baseline, recall at rank 10 (R@10) increases from 82.7\% to 87.2\% for audio-to-text retrieval, and from 81.3\% to 83.2\% for text-to-audio retrieval.

\section{PairMix}
\label{sec:format}

In this section, we introduce PairMix, the first audio-text paired augmentation, which is a simple but effective augmentation method that generates audio-text pairs. Previous vision-language augmentation \cite{hao2023mixgen} simply mixes randomly selected raw images and concatenates their corresponding texts. We adapt this approach to the audio-language domain to generate audio-text pairs while preserving the multi-modal relationship.

PairMix first randomly selects $N$ audio-text pairs $\{(a_i, t_i)\}_{i=1}^{N}$ containing audio $a_i$ and text $t_i$ and then mixes each modality separately to generate the new audio-text pair. Unlike vision-language MixGen which simply implements mixup for images, PairMix leverages the characteristic of audio modality by probabilistically implementing either waveform-level or mel spectrogram-level audio mixup to generate new mel spectrogram $\hat{s}$. Finally, PairMix generates a new audio-text pair $(\hat{s}, \hat{t})$ by concatenating text. It is formally written as:
\begin{equation} \label{PairMix}
\begin{split}
&\hat{s}_w = M(\sum^N_{i=1}\lambda_i a_i) \\
&\hat{s}_m = \sum^N_{i=1}\lambda_i M(a_i) \\
&\hat{s} = \gamma \hat{s}_w + (1-\gamma) \hat{s}_m\\
&\hat{t} = \text{Concat}(\{t_i\}_{i=1}^N),
\end{split}
\end{equation}
where $\lambda_i\in[0,1]$ for $i=1,...,N$ is a hyperparameter such that $\sum_{i=1}^N \lambda_i=1$, $M(\cdot)$ is a mel spectrogram transformation from waveform, $\gamma$ is randomly sampled from Bernoulli distribution with probability 0.5, and Concat$(\cdot)$ is a concatenation operator which concatenates text inputs. $\hat{s}_w$ and $\hat{s}_w$ denote mel spectrogram which is generated from waveform-level mixup and mel spectrogram-level mixup, respectively. Probabilistically applying two types of mixup allows us to generate more diverse samples, and we observe that this randomness contributes to stable improvements over hyperparameters as shown in Figure \ref{fig:test2}. Since PairMix can be easily applied with any other uni-modal augmentations, it is possible to drastically scale up the size of the dataset with a simple variant.

\section{Multi-level Test-time Augmentation}
\label{sec:pagestyle}

Test-time augmentation (TTA) contributes to generalizing models by making multiple predictions from augmented inputs and averaging the predictions. TTA has no additional training cost because TTA is only implemented at test-time as opposed to traditional augmentations which mainly focus on train-time. In this paper, we discuss the effects of conventional TTA for audio-language learning and present the generalized TTA method. As shown in Figure \ref{fig:tta}, conventional TTA only averages outputs from augmented inputs. Although this can contribute to performance, conventional TTA has limited performance improvement bound as shown in Figure \ref{fig:test2}.
To fully leverage TTA, we approach this problem with the view of the place of aggregation, unlike previous research \cite{kim2020learning, shanmugam2021better} that tends to focus on a selection of augmentation types and a method of aggregation. We present middle-level TTA (Mid-TTA) that averages intermediate representations and outputs a single prediction. Finally, we propose a novel multi-level TTA (Multi-TTA) methodology by generalizing conventional TTA and Mid-TTA. Unlike a single layer-based conventional TTA, Multi-TTA selects multiple layers for TTA so that augmentations are aggregated at various layers as shown in Figure \ref{fig:tta} (c).

We define a Multi-TTA strategy $S=(\{P_h\}_{h=1}^H,\tau)$ given a model $f$ containing $H$ layers where $\tau$ is the total number of input augmentations as follows. Let $P_h$ be a partition of $\{ 1,...,|P_{h-1}|\}$. Each element of $P_h$ contains indices of the previous layer's outputs that should be aggregated together. We define $P_h[i]$ as an $i$-th element of $P_h$ for simplicity. For clarity, $|P_h|$ decides the number of $h$-th layer's outputs, and $|P_h[i]|$ decides the number of required $h$-th layer's inputs for an $i$-th output of $h$-th layer. A strategy $S$ satisfies $|P_H|=1$ to output a single prediction and $\tau=\Pi_{h=1}^H|P_h|$. We define an $i$-th output of a $h$-th layer $o_{h,i}$ for $i=1,...,|P_h|$ as follows:
\begin{equation} \label{TTA}
o_{h,i}=\dfrac{1}{|P_h[i]|}\sum_{j\in P_h[i]}f_h(o_{h-1,j}).
\end{equation}
From the notations defined above, conventional TTA becomes a special case of Multi-TTA when $|P_h|=\tau$ for $1\leq h\leq H-1$ and $|P_H|=1$. Mid-TTA is also a special case of Multi-TTA when $|P_h|=\tau$ for $1\leq h\leq h'$ and $|P_h|=1$ for $h>h'$.

\begin{table*}[t]
\caption{Evaluation of the model performance in automated audio captioning.}
\label{table:aac}
\centering{}
\begin{tabular}{l|ccccccccc}
\hline
    Method  & BLEU$_1$ & BLEU$_2$ & BLEU$_3$ & BLEU$_4$ & METEOR & ROUGE$_L$ & CIDEr & SPICE & SPIDEr\\ 
   \hline
   Baseline \cite{mei2021audio} & 65.4 & 47.5 & 33.6 & 23.4 & 22.4 & 47.1 & 63.5 & 16.8 & 40.2\\   + PairMix & 69.3 & 52.9 & 38.9 & 28.3 & 24.1 & 49.9 & 75.5 & 17.7 & 46.6\\
   + Multi-TTA & \textbf{70.0} & \textbf{53.4} & \textbf{39.5} & \textbf{28.9} & \textbf{24.2} & \textbf{50.2} & \textbf{76.9} & \textbf{18.1} & \textbf{47.5}\\
   \hline

\end{tabular}
\end{table*}

\begin{table*}[t]
\caption{Evaluation of the model performance in audio-text retrieval.}
\label{table:retrieval}
\centering{}
\begin{tabular}{l|ccccc|ccccc}
\hline
    \multirow{2}{*}{Method} & \multicolumn{5}{c|}{text-to-audio} & \multicolumn{5}{c}{audio-to-text}\\ 
    \cline{2 - 11}
      & R@1 & R@5 & R@10 & R@50 & meanR & R@1 & R@5 & R@10 & R@50 & meanR \\ 
    \hline
    Baseline \cite{Mei2022-qx} & 33.0 & 67.9 & 81.3 & 96.5 & 10.0 & 37.9 & 71.0 & 82.7 & 97.4 & 8.8 \\
    + PairMix & 34.4 & 69.6 & 83.1 & \textbf{97.5} & 8.2 & 40.0 & 73.1 & 86.8 & 97.3 & 6.6 \\
    + Multi-TTA & \textbf{34.7} & \textbf{70.3} & \textbf{83.2} & \textbf{97.5} & \textbf{8.0} & \textbf{40.2} & \textbf{74.0} & \textbf{87.2} & \textbf{97.6} & \textbf{6.3} \\
    \hline
\end{tabular}
\end{table*}

\begin{table}[t]
\caption{Comparison between uni-modal and multi-modal data augmentation. In modality, A means Audio augmentations and T means text augmentations. Augmentation All denotes using every audio augmentation with PairMix.}
\label{table:unimodal}
\centering{}
\begin{tabular}{l|c|c|cc}
\hline
    \multirow{2}{*}{Augmentation} & \multirow{2}{*}{Modality} & \multirow{2}{*}{SPIDEr} & \text{T}$\rightarrow$ \text{A} & \text{A}$\rightarrow$ \text{T}\\
    & & & R@10 & R@10 \\ 
    \hline
    - & - & 41.6 & 78.9 & 81.9 \\
    Noise & A & 40.9 & 79.1 & 81.1 \\
    Reverb & A & 41.1 & 80.4 & 81.0\\
    SpecAugment \cite{park2019specaugment} & A & 40.2 & 81.3 & 82.7\\
    EDA \cite{wei2019eda} & T & - & 79.3 & 81.3 \\
    PairMix & AT & 43.0 & 82.2 & 84.9 \\ 
    \hline
    \hline
    All & AT & 46.6 & 83.1 & 86.8 \\ 
    \hline
\end{tabular}
\end{table}

\section{Experiments}

\subsection{Dataset}
AudioCaps is the largest audio-text paired dataset containing approximately 46K audio clips of 10 seconds extracted from AudioSet \cite{gemmeke2017audio} and their corresponding text descriptions. We investigate the capability of our proposed methods on AudioCaps for automated audio captioning and audio-text retrieval. We follow the train, validation, and test splits stated in \cite{kim2019audiocaps} and reproduce the baselines without unavailable audio clips.

\subsection{Audio-Language Learning}
\subsubsection{Automated Audio Captioning}

Audio Captioning Transformer (ACT) \cite{mei2021audio} has an encoder-decoder structure from Transformer \cite{vaswani2017attention}. The encoder block of ACT is initialized with DeiT \cite{touvron2021training} which was trained for image classification tasks and then pre-trained on an audio tagging task using AudioSet. We utilize the pre-trained ACT encoder\footnote{\href{https://github.com/XinhaoMei/ACT}{https://github.com/XinhaoMei/ACT}} and randomly initialize the ACT-m decoder for a baseline of AAC, resulting in 108M parameters.
For evaluation, we employ BLEU \cite{papineni2002bleu}, METEOR \cite{banerjee2005meteor}, and ROUGE-L \cite{lin2004rouge} from machine translation metrics. We also report CIDEr \cite{vedantam2015cider}, SPICE \cite{anderson2016spice}, and SPIDEr \cite{liu2017improved}. Captions are decoded with beam search up to size $3$.

\begin{table}[t]
\caption{Experiment with different augmentation design choices for audio mixing in multi-modal augmentation.}
\label{table:design-choice}
\centering{}
\begin{tabular}{l|c|cc}
    \hline
    \multirow{2}{*}{Design} & \multirow{2}{*}{SPIDEr} & \text{T}$\rightarrow$ \text{A} & \text{A}$\rightarrow$ \text{T}\\
    & & R@10 & R@10 \\ 
   \hline
   concat & 45.0 & 81.9 & 80.8 \\
   waveform-only & 43.7 & 80.5 & 84.1\\
   spectrogram-only & 44.6 & 82.3 & 83.3\\
   PairMix & 45.8 & 82.9 & 84.7\\ 
   \hline
\end{tabular}
\end{table}

\subsubsection{Audio-Text Retrieval}

Pre-trained encoders are employed and trained using contrastive loss for audio-text retrieval, following the previous study \cite{Mei2022-qx}. We choose PANNs \cite{kong2020panns} as the audio encoder, which has ResNet38 architecture trained on the audio tagging task using AudioSet. For the text encoder, we initialize our model with pre-trained BERT \cite{devlin2018bert} and extract a text representation using the [CLS] token. The final model has 187M trainable parameters. Retrieval models are evaluated by Recall at rank k (R@k) and mean rank (meanR).

\subsection{Experimental Setup}
We experiment PairMix with different ratios $K\in \{0.125, 0.25, 0.5, 0.6\}$ where $K$ is a ratio of \#(generated samples) in a mini-batch to a size of mini-batch. Also, we compare a fixed mixup ratio $\lambda=0.5$ to $\lambda\sim$ Beta$(0.1, 0.1)$. As models with $K = 0.25$ and $\lambda\sim$ Beta$(0.1, 0.1)$ show high and stable performance, we use this setting for later experiments. 

Additionally, we explore other design choices such as audio concatenation, waveform-level-only mixup, and mel spectrogram-level-only mixup for mixing audio in PairMix. We mix two audio-text pairs and load additional audio-text pairs only for PairMix to avoid duplication within a mini-batch.
To validate the effectiveness and compatibility of PairMix, we compared PairMix with other uni-modal augmentations. Gaussian noise and reverberation are employed as waveform-level audio augmentation, with a probability of 0.5. SpecAugment \cite{park2019specaugment}, which is already applied in the baseline model, was applied as a spectrogram-level audio augmentation. For text augmentation, easy data augmentation (EDA) \cite{wei2019eda} was used. Note that text augmentation is not applied in audio captioning as it can harm the ground-truth captions. 

We validate Multi-TTA strategies where TTA operation is implemented at an intermediate layer $h_E$ and an output layer $H$. We set $h_E$ as the last layer of the encoder for AAC and the last layer of the PANN audio encoder for audio-text retrieval. We compare the results of conventional TTA, Mid-TTA, and Multi-TTA with $\tau\in\{10, 25, 50, 100\}$. For Multi-TTA, we employ the strategy $S$ satisfying $(\tau,|P_{h_E}[i],|P_H[j]|)\in\{(10, 2, 5), (25, 5, 5), (50, 5, 10),\\(100, 5, 20)\}$ for $\forall i, j$.
As we already mentioned, $|P_l[k]|$ denotes the number of required $l$-th layer's inputs for a single $k$-th output of layer $l$. Note that we set $|P_l[k]|$ for $\forall k$ as identical value for simplicity of experiments and $S$ satisfies $\tau=|P_{h_E}[i]*|P_H[j]|$.
We stabilize a prediction by averaging the output at which no augmentation is applied and the output of TTA. Types of data augmentation on test-time are consistent with train-time. Multi-modal augmentations are not included because generated samples are needless for evaluation. In test-time, we halve the maximum width and height of SpecAugment masking, reverberation degree, and the probability for applying audio augmentations compared to train-time.

All models are trained for 30 epochs using AdamW optimizer of learning rate 10$^{-4}$ with the weight decay of 10$^{-6}$. SpecAugment is applied to every model. Other hyperparameters follow \cite{mei2021audio} for AAC and \cite{Mei2022-qx} for audio-text retrieval. Single NVIDIA GeForce RTX 3090 was used to train every model, taking 10 hours and 9 hours to train the AAC and audio-text retrieval models, respectively.

\begin{figure}
    \centering
    \includegraphics[width=1\columnwidth]{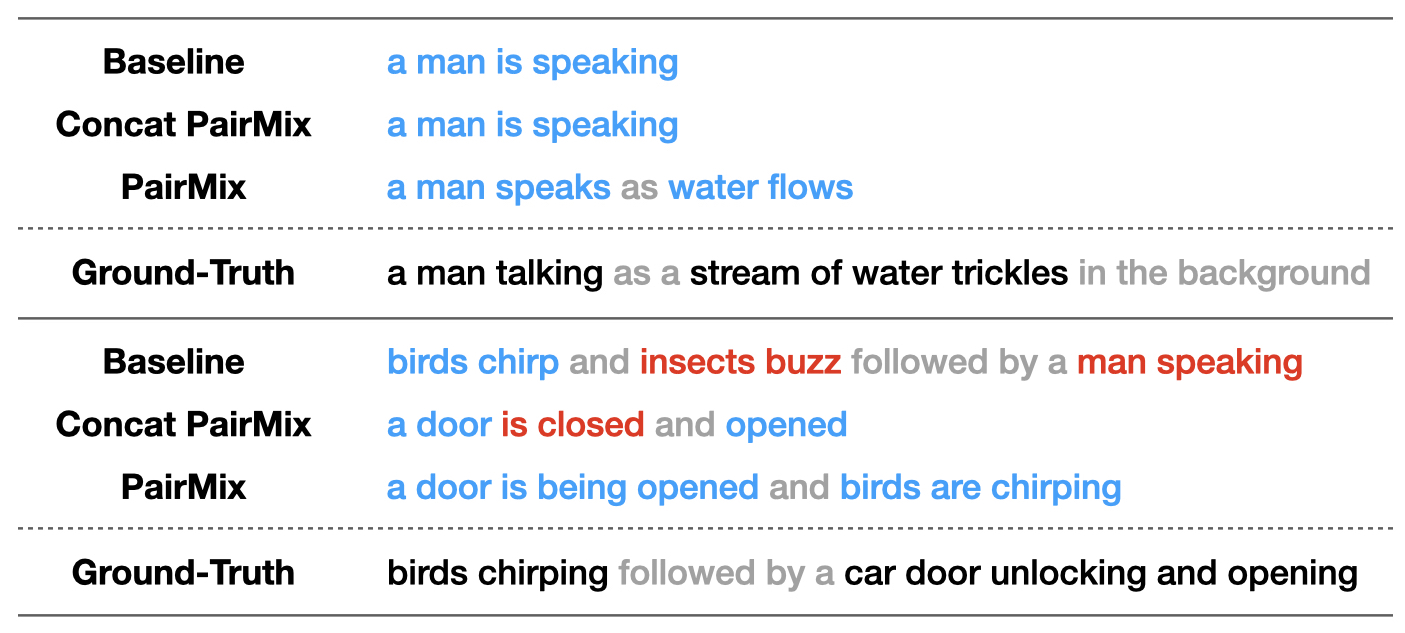}
    \caption{Examples of generated captions in automated audio captioning. PairMix correctly captures the overlapping events in a single audio clip, unlike the others.}
    \label{fig:fig_sample}
\end{figure}

\section{Results}
\label{sec:majhead}
Table \ref{table:aac} and \ref{table:retrieval} show an overview of the result. In AAC, using PairMix with the uni-modal audio augmentations achieves 46.6 SPIDEr, which outperforms the previous state-of-the-art performance with a 5x smaller model. In audio-text retrieval, applying PairMix increases R@10 to 86.8\% for audio-to-text retrieval and 83.1\% for text-to-audio retrieval. Furthermore, attaching Multi-TTA to AAC achieves 47.5 SPIDEr, which is an +18.2\% relative to the baseline. Adding Multi-TTA gives additional gain to audio retrieval as well, R@10 of audio-to-text retrieval and text-to-audio retrieval is 87.2\% and 83.2\%, respectively. 

\subsection{PairMix}
In Table \ref{table:unimodal}, we compare the proposed PairMix to existing uni-modal augmentations. As shown in the table, PairMix performs better than all the other uni-modal augmentations in both AAC and audio-text retrieval. The superiority of our PairMix demonstrates that paired audio-text data augmentation is critical to audio-language learning.  Furthermore, applying PairMix with uni-modal augmentations leads to additional improvement. 

Also, we explore other design choices of PairMix. Concat PairMix concatenates audio clips instead of mixing, and waveform PairMix and spectrogram PairMix only augments at waveform- and mel spectrogram-level, respectively. Table \ref{table:design-choice} demonstrates that our PairMix with probabilistic waveform- and spectrogram-level audio mixup is the best strategy. We investigate the generated captions of the baseline, PairMix, and concat PairMix to illustrate the difference. In failure cases, the baseline and concat PairMix miss or generate inaccurate explanations as shown in Figure \ref{fig:fig_sample}. However, PairMix generates robust captions with samples containing overlapped events.

\begin{figure}
  \centering
  \includegraphics[width=1\linewidth]{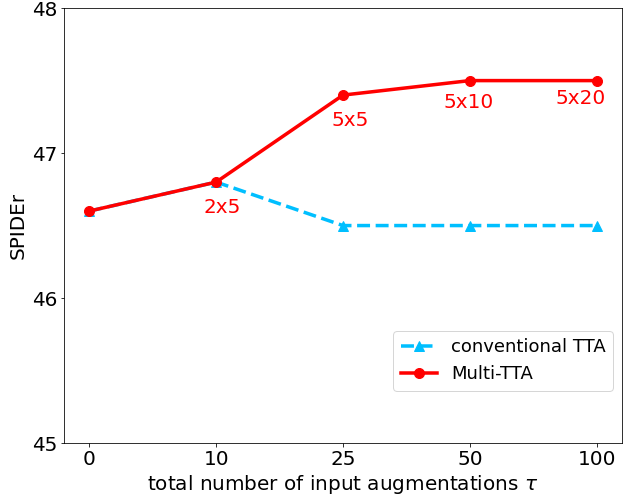}
  \caption{Experiment on TTA with various $\tau$ in automated audio captioning. Multi-TTA strategies are describes in the form of $|P_{h_E}[i]|\times|P_H[j]|$. Overall, Multi-TTA surpasses conventional TTA.}
  \label{fig:test2}
\end{figure}

\subsection{Multi-Level Test-Time Augmentation}
Figure \ref{fig:test2} demonstrates the relationship between the number of input augmentations and the output performance. Only Multi-TTA enhances as the number of input augmentations $\tau$ increases, even though they seem to be similar with a small number of augmentations. This indicates that aggregating augmentations in multiple layers enhances the bounded performance of conventional single-layer TTA, allowing the model to fully use the benefit of augmented samples. As a result, applying Multi-TTA to model with $\tau=100$, $|P_{h_E}[i]|=5$, and $|P_H[j]|=20$ successfully improves every evaluation metric in both AAC and audio-text retrieval.

There could be other hyperparameters to change like the selection of middle layers. Especially, unlike AAC where we can easily select intermediate layers to apply TTA as it has an encoder-decoder structure, audio-text retrieval models have more freedom to select any layers inside the encoder. Future research on automated layer search could further improve current results.

\section{Conclusions}
\label{sec:subhead}

We demonstrate that appropriate augmentations improve audio-language learning performance. Specifically, experimental results show that PairMix simply generates new audio-text pairs effectively. Also, we generalize TTA and propose Multi-TTA which enhances the efficiency of augmentations. Our proposed methods surpass the state-of-the-art method in AAC and show competitive results in audio-text retrieval. Since PairMix and Multi-TTA are flexible, they can be simply incorporated into other architectures.

\bibliographystyle{IEEEtran}
\bibliography{mybib}

\end{document}